\documentclass[final,3p,times]{elsarticle}
\usepackage[usenames,dvipsnames]{color}
\usepackage{psfrag}

\usepackage{amssymb}
\usepackage{amsthm}
\usepackage{amsmath}
\usepackage{graphicx}
\usepackage{float}
\usepackage{subfigure}
\usepackage{lineno}

\begin{document}

\begin{frontmatter}

  \title{Explicit filtering for large eddy simulation as use of a
    spectral buffer}

\author[]{Joseph Mathew\corref{mycorrespondingauthor}}

\cortext[mycorrespondingauthor]{Corresponding author}
\ead{joseph@aero.iisc.ernet.in}

\address{Department of Aerospace Engineering, Indian Institute of
  Science, Bangalore 560012, India} 

\begin{abstract}  
  The explicit filtering method for large eddy simulation (LES), which
  comprises integration of the governing equations without any added
  terms for sub-grid-scale modeling, and the application of a low-pass
  filter to transported fields, is discussed.  The shapes of filter
  response functions of numerical schemes for spatial derivatives and
  the explicit filter that have been used for several LES are
  examined.  Generally, these are flat (no filtering) over a range of
  low wavenumbers, and then fall off over a small range of the highest
  represented wavenumbers.  It is argued that this high wavenumber
  part can be viewed as a spectral buffer analogous to physical buffer
  (or sponge) zones used near outflow boundaries.  The monotonic
  convergence of this approach to a direct numerical simulation, and
  the shifting of the spectral buffer to larger wavenumbers as the
  represented spectral range is increased, without altering the low
  wavenumber part of solutions, is demonstrated with LES of two sample
  flows.  Connections to other widely used methods---the Smagorinsky
  model, MILES and another ILES---are also explained.
   
\end{abstract}


\end{frontmatter}
\modulolinenumbers[5]

\renewcommand{\thefootnote}{\fnsymbol{footnote}}

\section{Introduction}
\label{Introduction}

Large eddy simulation (LES) connotes a numerical simulation that is
restricted to a range of the largest scales of a turbulent flow.  The
approximation is useful when the solution is qualitatively correct,
and typical quantities of interest, such as flow statistics, are
obtained with acceptable accuracy.  The error is acceptable because of
the great reduction in computing effort (grid size, computation time)
compared with that required for a solution that contains all
dynamically significant scales.  There have been many reports of LES
by various techniques that have been developed to overcome shortfalls
of various approaches.  A comprehensive presentation of techniques and
sample results can be found in \citet{Sagaut2006}.  Much of the effort
had been directed at finding the best sub-grid-scale (SGS) model which
must capture the effect of the omitted small scales on the computed
large scales.  It cannot be said that any one approach has emerged as
a sole best method.  Instead, practitioners adopt a particular method
(numerical scheme and SGS model) which has proved successful for their
studies, and whose requirements for each class of flows (free shear
flows, wall-bounded flows) are known from experience.  This paper
discusses some new understanding of an explicit filtering approach
introduced in \citet{Mathew2003}, and connections to other popular
methods.  In the literature, explicit filtering can also refer to
filtering nonlinear terms alone, which has been examined by
\citet{Lund2003}.  It was devised to control numerical error and not
as an SGS model, unlike the model discussed here.

The derivation of the explicit filtering SGS model is described below.
Next, it's application to turbulent flows with two kinds of numerical
scheme-filter pairs is discussed.  The shapes of the filter response
functions motivates the idea of a spectral buffer.  Two examples show
the presence of this buffer at different grid resolutions, filter
cut-offs, and Reynolds numbers.  Connections to other SGS models are
discussed in \S~\ref{othersgs}.

\pagebreak
\subsection{Approximate deconvolution model}

The explicit filtering method of \citet{Mathew2003} was derived
from the approximate deconvolution model (ADM)~\cite{Stolz1999}.  Salient
aspects will be summarized below for completeness.  Consider the
one-dimensional evolution equation for $u(x,t)$,
\begin{equation}
\frac{\partial u}{\partial t} + \frac{\partial }{\partial x}f(u) =0,
\label{1d}
\end{equation}
where $f(u)$ is a nonlinear function.  An LES can be interpreted as
either obtaining an approximation $\bar{u}(x,t)$ that contains a large
scale part of $u(x,t)$, or the approximation that can be obtained
on a coarse grid which implicitly limits the range of wavenumbers in
the solution.  Then $\bar{u} = G * u = \int G(x-x')\; u(x')\; {\rm
  d}x$, where $G$ is a low pass filter.  The evolution equation for
$\bar{u}(x,t)$ is obtained by applying the filter to eqn.~(\ref{1d})
to get
\begin{equation} \label{1dG}
\frac{\partial \bar{u}}{\partial t} + G * \frac{\partial
}{\partial x}  f(u) =0. 
\end{equation}
Equation~(\ref{1dG}) can be written in the form of the original
equation with a remainder ${\cal R}$ 
\begin{equation} \label{1dles}
\frac{\partial \bar{u}}{\partial t} + \frac{\partial
}{\partial x}  f(\bar{u}) =  {\cal R},
\end{equation}
$$
{\cal R} =  \frac{\partial f(\bar{u})}{\partial x}
-  G * \frac{\partial f(u)}{\partial x}.
$$
${\cal R} \equiv 0$ unless $f(u)$ is nonlinear.  Since $u$ is not
known when solving for $\bar{u}$, ${\cal R}$ must be replaced with a
model ${\cal R}_m(\bar{u})$ for closure.  In ADM~\cite{Stolz1999}
\begin{equation} \label{radm}
{\cal R}_m =  \frac{\partial f(\bar{u})}{\partial x}
-  G * \frac{\partial f(u^*)}{\partial x},
\end{equation}
where $u^*(x,t) = Q* \bar{u}$ is an approximation to $u(x,t)$ obtained
by deconvolution of the filtered variable $\bar{u}$.  The equation
solved when using ADM is
\begin{equation} \label{adm}
\frac{\partial \bar{u}}{\partial t} + G * \frac{\partial
  f(Q*\bar{u})}{\partial x} =0. 
\end{equation}
During the early development of ADM, when the method was applied to
different types of problems, filters were obtained from implicit
formulas (Pad\'e-type, $G(\alpha)$ with different values of the filter
parameter $\alpha$\footnote[2]{$\alpha=-0.2$ in ref.~\cite{Adams1999},
  $\alpha=0.25$ in ref.~\cite{Stolz1999}.}) and from explicit
formulas~\cite{Stolz2001a}.  The deconvolution $Q*\bar{u}$ was
performed by applying the filter $G$ several times as per an expansion
of the operator $Q$ in terms of $G$.  In all cases, excellent results
for LES were presented.  Around the same time, \citet{Geurts1997}
examined a similar de-filtering, taking the primary filter to be a
top-hat function in physical space, and inversion to be exact for
polynomials.  It was not understood whether there was a best
convolution-deconvolution pair $G$ and $Q$, or that all pairs that
satisfied some property would be suitable. As will be shown below, the
solution depends on the effective filter $E = Q*G$ rather than on the
constituent filter and deconvolution operators themselves.

\subsection{Explicit filtering implementation of ADM}

The implementation of ADM by solving eqn.~(\ref{adm}) involves the
following steps to obtain $\bar{u}(x,t^{n+1})$ at the $n+1$ timestep
given $\bar{u}(x,t^n)$.
\begin{enumerate}
\item{Deconvolution: $u^* = Q * \bar{u}(x,t^n)$}
\item{Integration of eqn.~(\ref{adm}): $\bar{u}(x,t^{n}) \rightarrow
    \bar{u}(x,t^{n+1})$}
\end{enumerate}
This integration step by the Euler forward formula is
$$
\bar{u}(x,t^{n+1}) = \bar{u}(x,t^{n})
- \Delta t\, G * \frac{\partial f(u^*)}{\partial x},
$$
and can be evaluated in two steps
\begin{align}
\label{intstep}a(x) &=  u^*(x,t^n)  -\Delta t \frac{\partial
  f(u^*)}{\partial x} \\ 
\label{fstep} \bar{u}(x,t^{n+1}) &=  G * a(x) + \left [\bar{u}(x,t^{n}) -
  G*u^*(x,t^n) \right ]
\end{align}
The quantity within the square brackets in eqn.~(\ref{fstep}) is small,
and can be neglected, because the essential requirement for the
deconvolved field is that $u^* \approx u$ over a range of large
scales; then $ \bar{u} = G *u \approx G*u^*$.  The intermediate field
$a(x)$ can also be written as $u^*(x, t^{n+1})$.  This alternate
implementation comprises the following three steps:
\begin{enumerate}
\item{Deconvolution: $u^*(x,t^n) = Q * \bar{u}(x,t^n)$}
\item{Integration of eqn.~(\ref{1d}) with $u^*$ instead of $u$:
    $u^*(x,t^{n}) \rightarrow u^*(x,t^{n+1})$}\hfill  (applying
  equation~\ref{intstep}) 
\item{Filtering: $\bar{u}(x,t^{n+1}) =  G * u^*(x,t^{n+1})$}
\hfill  (applying equation~\ref{fstep}) 
\end{enumerate}
\citet{Mathew2003} observed that, in this latter form, the simulation
proceeds by repeating the following two steps: an integration of the
original evolution equation~(\ref{1d}) followed by a filtering and
deconvolution that can be combined into an (explicit) filtering of the
evolving field with a resultant filter $E=G*Q$ (step 3 of a time-step
is combined with step 1 of the following time-step).  Since $Q$ is an
approximate inverse of $G$ over a range of large scales, by
definition, $E \approx I$ over that range of large scales.  Beyond
that range, we would like $E$ filter out content.

In ADM (eqn.~\ref{adm}), a specific filter $G$ was assumed, a
procedure provided an approximate deconvolution operator $Q$, and,
thereby, an estimate $u^*$ as a structural model closure.  ADM
provided no guidelines on what $G$ ought to be, and different
operators $Q$ could be obtained for the same $G$.  Moreover, because
the operator $Q$ amplifies content within the represented range of
scales only, ADM offered no answer to those who expect an LES model to
account for the effect of small scales omitted from the
computation\footnote[3]{In \S~\ref{sgseffect} it is argued that such
  effects are not significant.}.  The derivation of the explicit
filtering method revealed a {\em principle} for LES: a structural SGS
model for LES is realized by integrating the governing equations
without adding any model terms and applying a flat, low-pass filter
to the transported variables after every time step; consistently,
discretization formulas for spatial operations must be high-resolution
ones that have little error over a range of large scales.

\section{Explicit filtering for LES of turbulent flow }

For LES of a turbulent flow, eqn.~(\ref{1d}) is replaced by the
Navier-Stokes equations.  For incompressible flow, the explicit
filtering method described above can be implemented by integrating the
momentum equation to obtain a velocity field, obtain the pressure
field, and correct the velocity field so that it is divergence free as
in a DNS.  Next, an explicit filter $E$ should be applied to this
velocity field, since that is the transported field.  For compressible
flow, there are two other transport equations for, say, density and
energy, and these fields would also be filtered.  Other intermediate
variables, like temperature or pressure, that appear in the equations
needn't be filtered.  Note also that it is not necessary to use
filters that commute with differentiation, since commutation is not
invoked at any stage in deriving this method.

When the momentum equation is written in terms of the LES field, the
remainder ${\cal R}$ is termed the SGS stress which requires an SGS
model.  SGS modeling can be classified broadly into structural and
functional models~\cite{Sagaut2006}.  A structural model is obtained
by replacing the full-spectrum fields in SGS terms with an
approximation obtained from the computable partial spectrum fields.
ADM is an example that replaces the velocity field ${\bf u}({\bf
  x},t)$ with the approximation ${\bf u}^*({\bf x},t)$ that has been
obtained by deconvolution of the LES field $\bar{\bf u}({\bf x},t)$.
A velocity estimation model is another kind where an approximation to
${\bf u}({\bf x},t)$ is obtained by an integration of the governing
equations (or an approximation that makes computations economical)
with a slightly larger spectral content and for short
durations~\cite{Domaradzki1997}.  An example of a functional model is
the Smagorinsky model.  A well-known feature of turbulent flow is the
net transfer of energy from larger scales to smaller scales.  In an
LES where only a large scale part of the flow is computed, this energy
transfer is prevented, causing a growth of high wavenumber content,
the appearance of wiggles in physical space, and the solution
diverges.  The Smagorinsky model dissipates spectral content at all
scales but increasingly at the highest wavenumbers.  The coefficient
controls the magnitude of dissipation, and when it is determined
dynamically from the evolving fields themselves~\cite{Germano1991},
the model has been found to be more useful.

A large number of studies have appeared that use the FDL3DI code which
combines an explicit filtering step with high-resolution, compact
difference schemes.  The method was first described in
\citet{Visbal1999}.  Examples of LES were reported
subsequently~\cite{Visbal2002, Rizzetta2008}.  Derivatives were
computed with 4th and 6th-order compact differences and conserved
variables were filtered with 8th and 10th-order Pad\'e filters; there
were no added SGS model terms in the equations that were solved.  This
approach is quite similar to those described above~\cite{Mathew2003,
  Chakravorty2010}.  The numerical scheme and the explicit filter have
flat response functions with a smooth fall-off near the high
wavenumber end.  \citet{Bogey2004} proposed high-order explicit
difference schemes (8, 10 and 12th-order) with optimized coefficients
that have good resolution characteristics like compact schemes.  They
also devised a selective filter with response functions similar to
that shown in Fig.~\ref{frfa} (see Fig. 3 in their
paper~\cite{Bogey2004}).  They have used this method for LES of round
jets and computed turbulence profiles and the radiated sound.  Marinc
\& Foysi~\cite{Marinc2012} used an optimized 6th-order finite
difference scheme for spatial derivatives and an optimized 10th-order
explicit filter for their LES for control of aeroacoustics of plane
jets.  Explicit filtered LES was applied successfully to a reacting
plane jet injected into a compressible channel
flow~\cite{Schaupp2012}.  Foysi \& Sarkar~\cite{Foysi2010} added a
dynamic Smagorinsky term and applied explicit filtering for their LES
of round jets in the manner of mixed models; filtering was considered
to provide a reconstruction and the Smagorinsky term to provide
stabilization.  As will be shown below, it has been possible to
perform LES of round jets without the need for any additional terms.

\subsection{Filter characteristics}
\begin{figure}[h]
\psfrag{x}{$k/k_{\rm max}$}
\psfrag{y}{Filter response functions}
\includegraphics[width=0.4\textwidth]{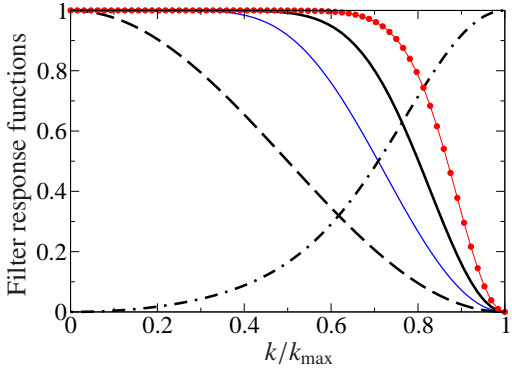}
\caption{Filter response functions associated with Pad\'e filter defined
  by eqn.~\ref{fpade}. -- -- --: $\hat{G}(k; \alpha=0)$,
  --~$\cdot$~--~$\cdot$~--: $(\hat{Q}_{ADM}(k)-1)/J; \alpha=0, J=5)$, {\bf
    ------}: $\hat{E}(k; \alpha=0)$, {\color{blue}------}: $\hat{E}(k;
  \alpha=-0.2)$, {\color{red}--$\bullet$--}: $\hat{E}(k; \alpha=0.2)$}
\label{frfa}
\end{figure}

Although derived as a structural model, the explicit filtering model
of \citet{Mathew2003} is a functional model. Before elaborating, let
us consider examples of $G$, $Q$ and $E=QG$.  Let $u_j$ denote values
of the function on a uniform grid of $N$ points $x_j = jh\;\;\;(j=0,1,
\ldots, N)$.  When combined with end-point formulas, the
following implicit formula provides a filtered field $\bar{u}_i$
\begin{equation}\label{fpade}
\alpha \bar{u}_{j-1} + \bar{u}_{j} + \alpha \bar{u}_{j+1}
= (\alpha + \frac{1}{2})\left (u_j + \frac{1}{2}(u_{j-1} +
  u_{j+1})\right ).
\end{equation}
The sole free parameter $\alpha$ controls the shape of the filter
response function.  We write $\bar{u} = G*u$.  For simplicity, suppose
$u$ to be periodic, with period $2\pi$.  Then, Fourier coefficients of
wavenumber $k$ are related as $\hat{\bar{u}}(k) =
\hat{G}(k)\,\hat{u}(k)$, where variables with carets are coefficients
of the appropriate Fourier series.  Figure~\ref{frfa} shows
$\hat{G}(k)$ with $\alpha=0$.  Note that there is significant
filtering at all wavenumbers.  Owing to the restriction of the
function $u$ to the grid of $N$ intervals, the spectral content of the
filtered field is restricted to $0 \le k \le k_{\rm max}$ and $k_{\rm
  max} = N/2$.  An exact inverse is not possible because content with
$k > N/2$ is not available.  An obvious approximate inverse is
$$
\hat{G}_N^{-1} = 
\begin{cases}
    1/\hat{G} &\;\;\;\; (0 \le k <N/2)\\
    0 &\;\;\;\; (k=N/2)
  \end{cases}
$$
Formally then, the deconvolution filter $Q=G_N^{-1}$.  The explicit
filter $\hat{E} = \hat{G}\hat{Q} =1\;\;(0 \le k < N/2)$ and vanishes
for $k=N/2$.  LES with these filters amounts to integrating the
transport equation without any explicit filtering.  The explicit
filtering SGS model is then inactive, will prove inadequate, and the
code will diverge, {\em unless} the filtering due to {\em other}
operations such as numerical differentiation provide the expected
functionality.

The approximate deconvolution operator proposed by \citet{Stolz1999}
is the truncated series
$$
Q_{\rm ADM} = \sum_{j=0}^J (I - G)^j
$$
where $I$ is the identity operator.  Figure~\ref{frfa} shows the
filter response function for $\hat{Q}_{\rm ADM} (\alpha=0)$ when 6
terms in the expansion are taken.  Also shown are the implied explicit
filters $\hat{E}=\hat{G}\hat{Q}_{\rm ADM}$ for $\alpha = -0.2, 0,
0.2$.  The explicit filter $E$ is flat (no filtering) over a range of
low wavenumbers and then falls smoothly to zero over a small part of
the highest represented wavenumbers.  As $\alpha$ increases the
cut-off wavenumber $k_{\rm cutoff}$ increases. A suitable definition
is of $k_{\rm cutoff}$ is that $\hat{E}(k_{\rm cutoff})=0.9$.

LES by explicit filtering with filters $E$, with characteristics
similar to those in Fig.~\ref{frfa}, have proved to be successful.
Several examples can be found in \citet{Rizzetta2008} of simulations
of many different kinds of flows with the AFRL code FDL3DI.  Another
common feature of these successful simulations is that the numerical
methods comprised either high-resolution, implicit (compact)
difference or very high-order, explicit difference formulas that also
provide high-resolution.  Of course, this is understandable:
difference formulas may be viewed as providing a spectrally-accurate
derivative combined with low-pass filtering.  The implied filter of a
symmetric, implicit difference formula or a high-order explicit
difference formula is nearly flat over a range of low wavenumbers and
then falls off to zero.  When such difference formulas are used the
governing equations are integrated accurately over a range of low
wavenumbers.

\begin{figure}[h]
\psfrag{x}{$k/k_{\rm max}$}
\psfrag{y}{Filter response functions}
\includegraphics[width=0.4\textwidth, trim=0 0 0 -25]{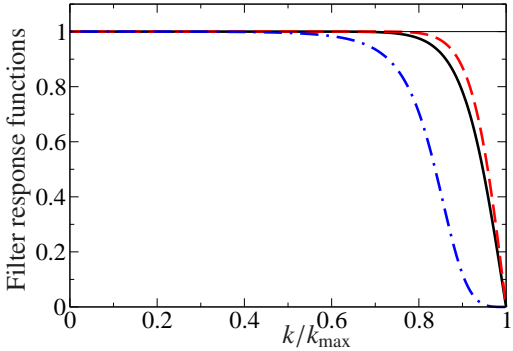}
\caption{Filtering of formulas designed by
  \citet{Chakravorty2010}.  ------: first derivative;
  {\color{red}-- -- --}: interpolation;
  {\color{blue}--~$\cdot$~--~$\cdot$~--}: explicit filter $E$.  }
\label{scfrf}
\end{figure}

\citet{Chakravorty2010} performed LES of incompressible and variable
density (low-Mach number) flows using compact difference formulas.
For the staggered grid algorithm, the needed interpolations were also
performed using high-resolution, implicit
formulas\cite{Chakravorty2004}.  Filtering characteristics of all
numerical procedures were designed to be flat over a range of low
wavenumbers and to fall off at high wavenumbers, like the explicit
filter $E$ shown in Fig.~\ref{frfa}.  Five-point stencils were taken
and constrained at two or three wavenumbers to design derivative and
interpolation formulas with a large range of wavenumbers over which
near-spectral accuracy would be obtained.  Truncation error was
4th-order for all formulas.  The explicit filter, defined on a 5-point
stencil, has one free parameter and is 4th-order.  The filter response
functions for the 1st derivative, interpolation and the explicit
filter are shown in Fig.\ref{scfrf}.  Note that there is a clear
separation between the cut-off wavenumbers of procedures in the
numerical scheme and the explicit filter.  The cut-off is at a smaller
wavenumber for the explicit filter ($k_{\rm cutoff} \approx 0.7k_{\rm
  max}$).  Below the cut-off, spectral accuracy is maintained.  A
similar relation between cut-offs of explicit filter and numerical
scheme was used in \citet{Mathew2003} for LES of supersonic channel
flow.

\begin{figure}[h]
\psfrag{x}{$k/k_{\rm max}$}
\psfrag{y}{Filter response functions}
\includegraphics[width=0.4\textwidth, trim=0 0 0 -25]{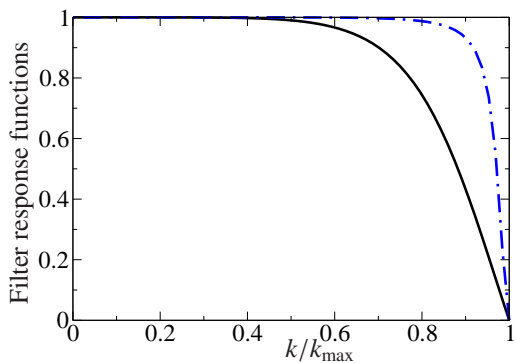}
\caption{Filtering of standard 6th-order compact difference formula
  (------) and 10th-order filter F10 with $\alpha=0.498$ 
  ({\color{blue}--~$\cdot$~--~$\cdot$~--}).}
\label{HT6frf}
\end{figure}

The earlier papers \cite{Mathew2003, Mathew2006} had advocated that
the cutoff wavenumber of the explicit filter be smaller than that of
the numerical method.  The operations in \citet{Chakravorty2010} have
also followed this principle.  However, it is not necessary because it
is the combination that is effective in a simulation.  For the LES of
\citet{Visbal2002}, when the 10th-order filter is applied with a high
value for the filter parameter of 0.49, the cut-off of the filter is
very close to the maximum wavenumber and clearly larger than that of
the derivative operations.  We have a similar experience when using
the MacCormack-type splitting scheme of \citet{Hixon2000}.
Figure~\ref{HT6frf} shows filtering characteristics of the standard
6th-order compact difference formula (eqn. (2.1.7) in
\citet{Lele1992}), and a 10th-order Pad\'e filter (filter F10 in table
IV of \citet{Visbal2002} with parameter $\alpha = 0.498$).  The
implied filter of the difference formula has a smaller cut-off than
the optimized schemes shown in Fig.~\ref{scfrf}.  So it has been
sufficient to use a filter with a higher cut-off than that of the
difference formula~\cite{Subramanian2013, Patel2015}.  This strategy
has been used for the round jet simulations discussed in \S \ref{sectrj}
below.

\subsection{Explicit filtering as employing a spectral buffer}

For simulations of unsteady, spatially developing flows, it is a
widespread practice to have a buffer zone between the outflow boundary
surface and the region of interest.  Within the buffer zone the
solution will not have the same accuracy as that in the region of
interest.  It may even be quite wrong.  If the grid spacing in the
buffer zone is increased aggressively, smaller scale motions that
convect into the buffer zone become damped numerically because they
can no longer be represented in coarse regions.  Damping can also be
effected by adding a term to the differential equations that takes the
solution to a smoother profile.  Then, at the outflow boundary, a
simple convective condition with a uniform convection velocity, or the
more detailed treatment based on characteristics~\cite{Thompson1987,
  Poinsot1992} can be applied.  In such cases there is an expectation
that any significant error due to the buffer zone treatment will be
restricted to the buffer zone, and to a smaller extent upstream of the
buffer, to about the thickness of the outgoing shear layers.

Our studies with the explicit filtering method indicate that the
result of filtering is analogous to using a buffer zone in spectral
space.  The buffer zone spans a small range of the largest represented
wavenumbers, and the corresponding range of represented high
frequencies.  Though these are small scale motions that are computed,
the filtering damps the energy contained, smoothly and increasingly
towards $k_{\rm max}$.  For all wavenumbers smaller than those of the
spectral buffer, all spatial operations are obtained with spectrally
accurate schemes.  The notion of a spectral buffer is suggested
strongly by the shape of the filter response functions of the
numerical schemes and the explicit filter.  In all computations
employing this approach, these functions have an essentially flat
portion over a significant range of represented wavenumbers and a
smooth fall-off.  For the two kinds of computations represented in
Figs.~\ref{scfrf} and \ref{HT6frf}, the spectral buffer spans the
approximate range $0.7 < k/k_{max} < 1$. In this range, the numerical
schemes have significant errors.  Errors in the solution that would
cause divergence begin to appear because the physical (and
mathematical) requirement of energy transfer to smaller scales ($k >
k_{max}$) is not met on an LES grid.  Explicit filtering with a
high-resolution filter meets one part of this requirement that the
energy be transferred out of this spectral range.  Just as one is not
concerned about what happens within and beyond a physical buffer
layer, one should not be concerned about what happens within and
beyond the spectral buffer.  However, it is only when the response
functions have nearly flat characteristics over a range of low
wavenumbers that there is a small spectral buffer zone.  When
low-order schemes are used, there is significant filtering even of low
wavenumber content, making the method less efficient: a relatively
finer grid is needed for the same level of accuracy over a fixed range
of large scales.  This is the functional modeling provided by explicit
filtering as an SGS model, though it was derived as a structural
model.

The term `spectral buffer' appears in Adams\cite{Adams2011} as the
range of represented scales that are not resolved by the numerical
method.  He explains that ADM was intended to `amplify scales in this
buffer' and a `relaxation term was introduced as a dissipative
mechanism on this buffer range.'  So the term was used to label the
range of scales where ADM procedures were to be active.  Deconvolution
amplifies scales in the resolved range as well.  In this paper, the
term spectral buffer is used as an analog of outflow buffers in
physical space, and not just as a name for a scale range where some
procedures have some effect.  It is then implied that the exact
treatment of the content in the spectral buffer is unimportant, rather
than fulfilling, say, a specific rate of absorption of kinetic energy,
and that an accurate treatment of larger scales is crucial.

\section{LES examples}\label{sectrj}

\begin{figure}
\psfrag{x}{$k\eta$}
\psfrag{y}{\hspace*{-13mm}$<\epsilon>^{-1/4} \nu^{-5/4}(k/\eta)^{5/3}
  <{\cal E}(k)>$}
\psfrag{a}{$32^3$}
\psfrag{b}{$48^3$}
\psfrag{c}{$64^3$}
\psfrag{d}{$192^3$}
\includegraphics[width=0.4\textwidth, trim=0 0 0 -25]{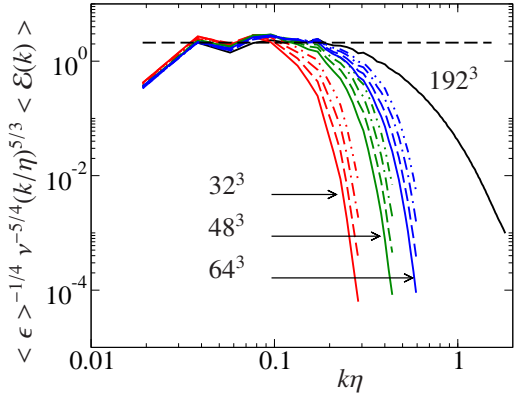}
\caption{Compensated energy spectrum from forced homogeneous,
  isotropic turbulence simulations of
  \citet{Chakravorty2010}.  LES grid sizes are $32^3$,
  $48^3$ and $64^3$, and DNS is with $192^3$ points.  For each group
  of LES there are 4 curves corresponding to simulations with 4 filter
  cutoff values.}
\label{hitEk}
\end{figure}

When there is such a clear separation between cut-off wavenumbers of
numerical scheme and explicit filter, it is possible to examine the
LES obtained as the cut-off wavenumber of the explicit filter alone is
changed.  With LES of supersonic channel flow \citet{Mathew2003}
have shown {\em monotonic} convergence towards the DNS solution as the
LES grid was refined, or as filter cut-off was increased.  An even
more compelling demonstration was obtained by \citet{Chakravorty2010}
by performing several LES and a DNS of forced homogeneous, isotropic
turbulence.  Figure~\ref{hitEk} shows the compensated energy spectrum
as a function of wavenumber $k$ scaled with the Kolmogorov length
scale $\eta$.  The DNS was on a grid of $192\times192\times192$
points.  LES were conducted on grids of $32^3$, $48^3$, and $64^3$
points, with four different values of cut-off wavenumbers of the
explicit filter.  Clearly, a) LES solutions converge monotonically to
the DNS with grid refinement or when filter cutoff is increased on a
given grid, and b) the changes are to high wavenumber content only.
As grid size increases there is essentially no change to low
wavenumber content even as the spectral range increases.  If one
compares any one of the LES, say, on the $32^3$ grid and with the
smallest filter cut-off, with the DNS, one infers that the explicit
filtering is providing a spectral buffer beyond scaled wavelength
$k\eta \approx 0.1$.  The fall-off of the spectrum is due to the
filter since the numerical differentiation is spectrally accurate over
a larger range.  On a given grid, as the filter cutoff wavenumber is
increased, the spectral buffer zone becomes thinner.  Solutions on
different grids demonstrate a complementary feature: on a finer grid,
the buffer zone has moved to a range of larger wavenumbers.
Significant errors remain confined to this buffer zone.  While it not
{\it a priori} evident that errors would be confined to wavenumbers
larger than the scheme cutoff, these results show no significant
contamination of smaller wavenumber content.

A second example is that of spatially developing, compressible round
jets at very high Reynolds numbers.  Simulation parameters are listed
in Table~\ref{rjpar}.  The Reynolds number Re = $U_0D/\nu$ is based on
jet diameter $D$ and its centerline velocity $U_0$ at the nozzle exit
plane.  The Mach number is $M = 0.9$.  A near top-hat velocity profile
with a tanh bounding shear layer was specified at the inflow plane.
Small-amplitude, random fluctuations were imposed on the shear layer
alone.  Simulations A and B are at Re = 11000, but for different
domain sizes---axial distances of $30D$ and $70D$, but with
approximately the same grid spacing.  \citet{Bogey2004} have simulated
a round jet for $75D$ at these conditions (Re = 11,000, $M = 0.9$).
Cases C and D are at the much higher Reynolds number of 1.1 million.
Spatial differences were obtained with a 6th-order compact scheme, and
a 10th-order filter was applied to conserved variables after every
time-step.  Filter response functions are those in Fig.~\ref{HT6frf}.
Time-stepping is with a 2nd-order, explicit, Runge-Kutta scheme.
Non-reflecting boundary conditions \citep{Poinsot1992} were applied at
the downstream and lateral boundary surfaces.  The numerical method
has been discussed in detail elsewhere~\citep{Patel2015}.  An
extensive discussion of the solution will also become
available~\citep{PatelPhD}.

\begin{figure}[h]
\subfigure[Re = 11,000, case A]{
\includegraphics[width=0.45\textwidth]{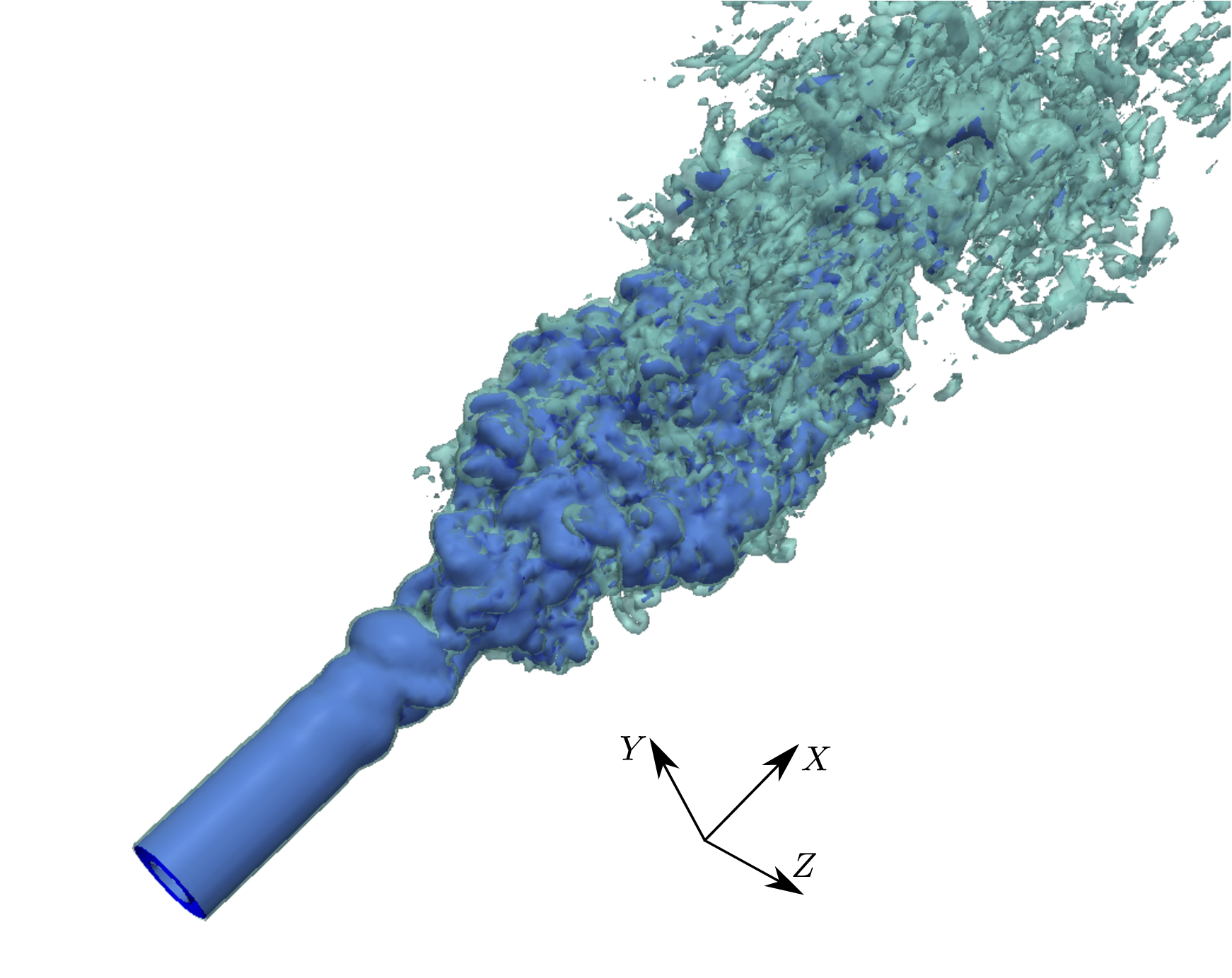}
}
\subfigure[Re = 1,100,000, case D]{
\includegraphics[width=0.45\textwidth]{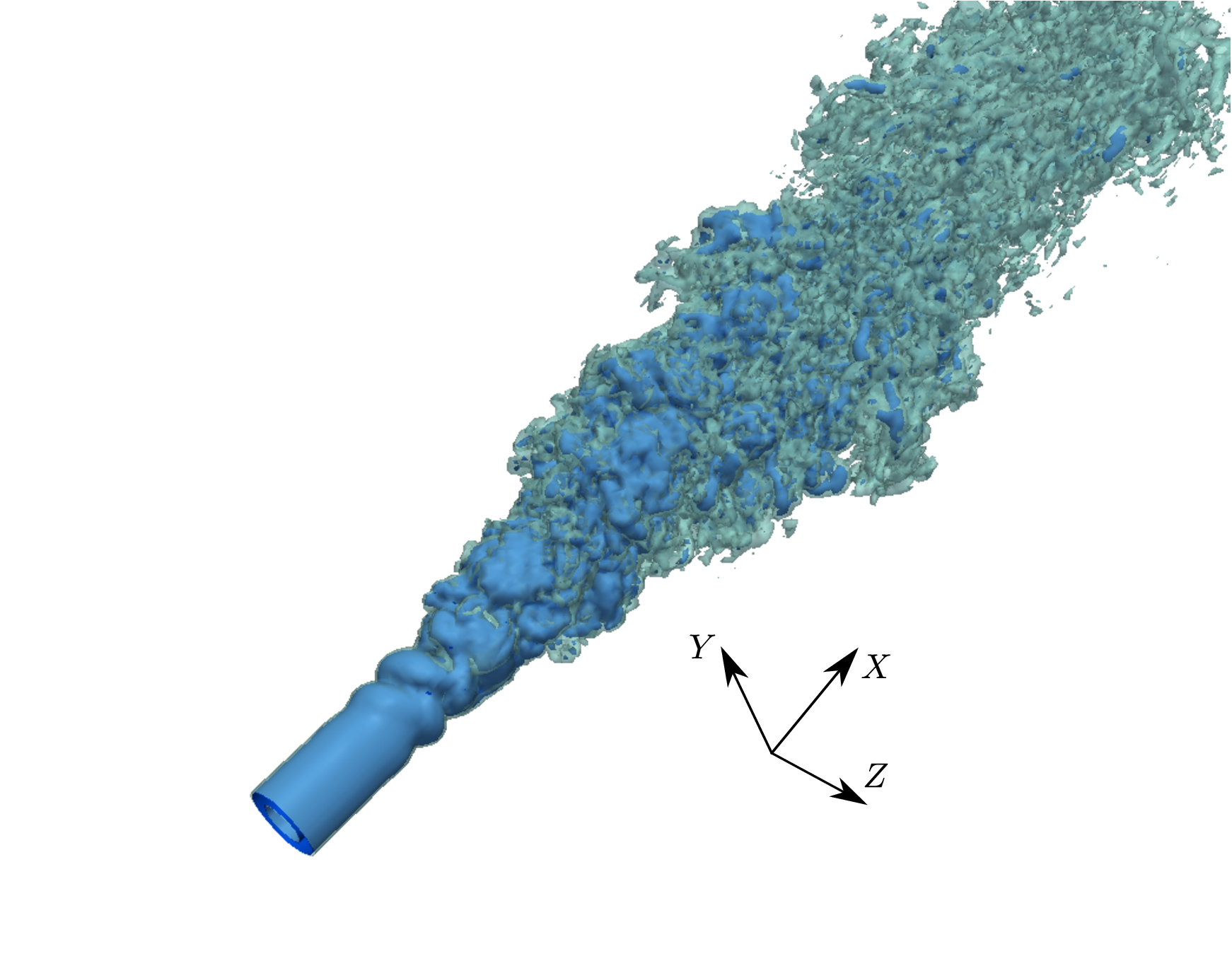}
}
\caption{Isosurfaces of vorticity magnitude $\omega D/U_0 = 7.8$,
  15.6.} 
\label{rj3d}
\end{figure}

The Cartesian reference frame used is shown in Fig.~\ref{rj3d}; the
$x$-axis was aligned with the jet and the origin is at the center of
the jet on the inflow plane.  The domain of interest is $0 < x < L_x$,
$-L_y/2 < y < L_y/2$, $-L_z/2 < z < L_z/2$.  The numbers of gridpoints
in this region are also given in Table~\ref{rjpar}.  Within this
region, the grids were stretched in lateral directions outside a
central square of side $1.5D$; grid spacing was increased in geometric
progression by 1\%.  For cases B and D, the grid was stretched axially
as well at 0.7\%.  The actual computational region was larger because
a buffer zone was used near the downstream and lateral boundaries
where the grid is stretched aggressively at 10\%.  There are
additional gridpoints in these buffer zones, 30 in the axial and 20 in
the lateral directions.  Fine scale structures disappear in the buffer
zone as they cannot be represented on the coarser grid.

\begin{table}\centering
\begin{tabular}{crccccc}
case & Re~~~~~ & $L_x/D$ & $L_y/D$, $L_z/D$ & $N_x$ & $N_y$, $N_x$
&$B_u$\\[2ex]
\hline\\
A& $1.1 \times 10^4$  & 30 & 10 & 263 & 253 &6.01\\
B& $1.1 \times 10^4$  & 70 & 40 & 426 & 473 &5.92 \\
C& $1.1 \times 10^6$  & 30 & 10 & 263 & 253 & 7.04\\
D& $1.1 \times 10^6$  & 30 & 10 & 406 & 387 & 6.03\\[1ex]
\hline
\end{tabular}
\caption{Round jet simulation parameters}
\label{rjpar}
\end{table}

\begin{figure}[h]
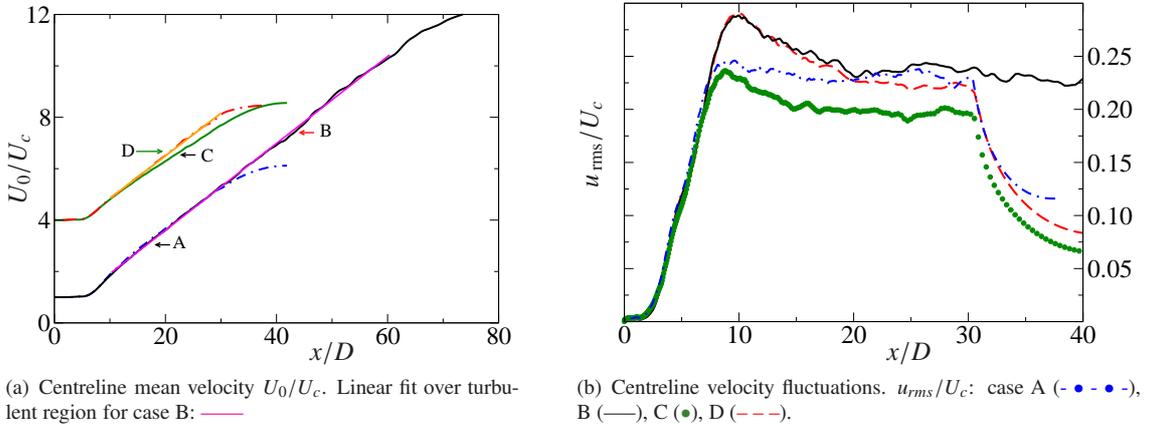
\centering
\vspace*{5mm}
\psfrag{y}{\hspace{-4mm}$U_0/U_c$}
\psfrag{x}{\hspace{5mm}$x/D$}
\subfigure[Centreline mean velocity $U_0/U_c$.  Linear fit over
turbulent region for case B:
{\color{magenta} ------}]{
\includegraphics[width=0.4\textwidth]{uc.eps}
\label{uc}
}\hfil
\subfigure[Centreline velocity fluctuations.
$u_{rms}/U_c$:~~case A ({\color{blue}- $\bullet$ - $\bullet$ -}), 
~~B (-----), C ({\color{OliveGreen} $\bullet$}), 
D ({\color{red} -- -- --}).
]{
\psfrag{x}{\hspace{5mm}$x/D$}
\psfrag{urms}{$u_{\,\rm rms}/U_c$}
\includegraphics[width=0.44\textwidth]{urms.eps}
\label{urms}
}
\caption{Streamwise development of jet.  In ($a$), curves for cases C
  and D have been vertically offset by 3 units for clarity.}
\label{ucurms}
\end{figure}

An impression of the scale range is conveyed in Fig.~\ref{rj3d} of
iso-surfaces of vorticity magnitude from cases A and D.  The larger
scale range at the higher Reynolds number is evident.  A sensitive
test of the correctness of these solutions is the development of the
inverse of the centerline velocity and the level of velocity
fluctuations.  The inverse of the centerline velocity is shown in
Fig~\ref{uc}.  For clarity, the curves for cases C and D have been
offset upward by 3 units.  The change in slope at $x/D \approx 10$ is
the beginning of the turbulent portion.  In all cases there is a clear
linear range following breakdown.  The curved portion near the
downstream end is the from the outflow buffer region.  Comparing cases
A and B, we observe that, if the domain is extended, the slopes of
these curves are essentially the same.  Linear fits for case B over
the portion $10 < x/D < 60$, and over $10 < x/D < 30$ for case D have
been included.  The reciprocal of the slope of the linear fits, $B_u$,
are listed in Table~\ref{rjpar}.  The values for cases A, B and D are
close to each other and agree closely with values from experiments
(5.8--6.06; Table 5.1 in Pope~\cite{Pope:2000}).  For the coarse grid
case C, the slope is significantly smaller.  Incidentally, the curves
for cases A and B also illustrate the correct effect of using a
physical buffer layer: The solution for A departs from the solution
for B only within its buffer layer.

Figure~\ref{urms} shows the development of velocity fluctuations along
the centerline, scaled with the local centerline mean velocity.  For
cases A, B and D, axial component $u_{\rm rms}/U_c$ tends to 0.24, as
in experiments~\citep{Panchapakesan:1993, Hussein:1994}.  Fluctuation
levels are smaller for the coarse grid case C, and is consistent with
the weaker decay rate of the centerline velocity (Fig.~\ref{uc}).  The
grid employed for case C is inadequate, but on refinement (case D), an
acceptable LES has been obtained.  Also, as in experiment, both
cross-stream components $v_{\rm rms}/U_c$ and $w_{\rm rms}/U_c$ were
found to tend to 0.18 (not shown here).

\begin{figure}[h]
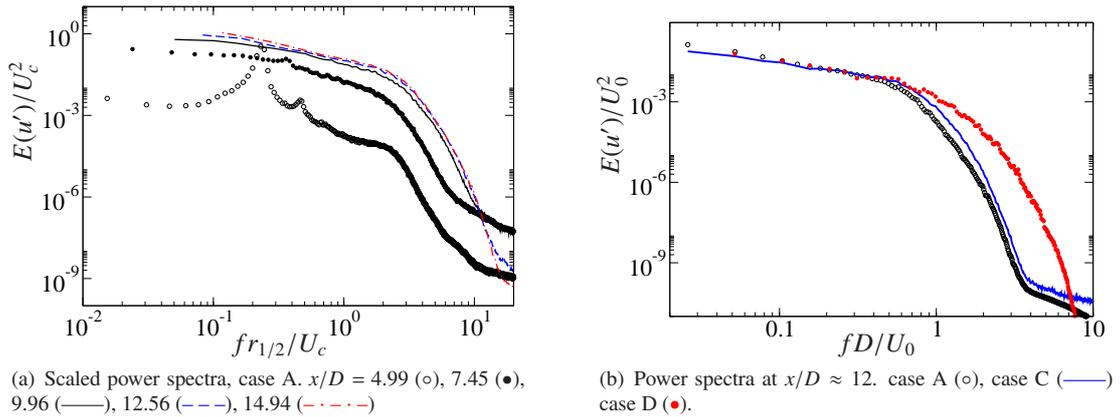
\centering
\vspace*{5mm}
\psfrag{x}{\hspace{-8mm}$fr_{1/2}/U_c$}
\psfrag{y}{$E(u')/U_c^2$}
\subfigure[Scaled power spectra, case A.
$x/D =$ 4.99 ($\circ$), 7.45 ($\bullet$), 9.96 (------), 12.56
({\color{blue}-- -- --}), 14.94 ({\color{red}-- $\cdot$ -- $\cdot$ --})
]{
\includegraphics[width=0.4\textwidth]{uspA.eps}
\label{uspA}
}\hfil
\subfigure[Power spectra at $x/D \approx 12 $. case A ($\circ$), 
case C ({\color{blue}------})
case D ({\color{red}$\bullet$}).
]{
\psfrag{x}{\hspace{-5mm}$fD/U_0$}
\psfrag{y}{$E(u')/U^2_0$}
\includegraphics[width=0.4\textwidth]{uspall.eps}
\label{uspall}
}
\caption{Power spectral density.}
\label{rjspec}
\end{figure}

Time series of velocity components at many stations on lines $y=0,
z=0$ (jet centerline) and $y=D/2, z=0$ (jet boundary shear layer at
inflow plane) were stored.  Frequency spectra were calculated using
the PWELCH function in MATLAB~8.6.  Figure~\ref{uspA} shows frequency
spectra of the streamwise velocity component, scaled with the local
mean centerline velocity, $E(u)/U_c^2$, at $x/D = 4.99$, 7.45, 9.96,
12.56, and 14.94 from case A.  Frequency $f$ has been scaled with the
jet half-radius $r_{1/2}$ and $U_c$.  Close to the inflow plane the
spectral range is small and solution field is fully represented and
accurately computed on the chosen grid.  Regular oscillations due to
the vortex rings upstream of breakdown (see Fig.~\ref{rj3d}) can be
observed as a low frequency peak in the spectrum at $x/D = 4.99$.
Downstream, this peak disappears as the flow breaks down to turbulence,
and the spectrum broadens.  Spectra collapse on these local scales to
show self-preserving development for $x/D \ge 9.96$.
Figure~\ref{uspall} shows spectra from cases A, C and D at $x/D
\approx 12 $.  When the Reynolds number alone is changed, the spectra
should extend to smaller frequencies as the inertial range extends,
while the content at low frequencies should remain approximately the
same (cases A and D).  Clearly, the simulations support this
expectation.  When the grid is refined, the spectrum extends as
smaller lengths and frequencies can be represented and computed
accurately (cases C and D).  Again, the changes are to the high
frequency end of the spectrum, while the low frequency part of the
solutions remains essentially the same.

The analogy between the spectral buffer of explicit filtering and
physical outflow buffer regions can be taken a little further.
Outflow buffer or sponge zone treatments are designed to continuously
suppress fluctuations until simple conditions applied to a smoothed
flow proves effective.  If the change is abrupt at the interface
between the region of interest and the buffer, the objective is not
realized and problems appear at the interface.  Filter response
functions of explicit filters used in all the cases cited here fall
off smoothly near the high wavenumber end as in Figs.~\ref{frfa},
\ref{scfrf} or, \ref{HT6frf}.  If the explicit filter had a sharp
cutoff at some $k_{\rm cutoff} < k_{\rm max}$, there would be energy
accumulation near $k_{\rm cutoff}$ and the computations would diverge.

\section{Other SGS models}\label{othersgs}
With this understanding we can consider the effects of some other
approaches.

The Smagorinsky SGS model is implemented as a term added to the
momentum equations by setting the SGS stress tensor
$$
\tau_{ij}^{\rm sgs} = \nu^{\rm sgs} S_{ij},
$$
where $S_{ij}$ is the strain rate tensor and $\nu^{\rm sgs}$ is the
eddy viscosity.  The effect of this term is to damp {\em all} scales.
Briefly, consider the 1-d equation (\ref{1dles}) for the `LES'
variable $\bar{u}$.  A Smagorinsky-type model for $\cal R$ would
appear as
\begin{equation} \label{1dlesSMG}
\frac{\partial \bar{u}}{\partial t} + \frac{\partial
}{\partial x}  f(\bar{u}) =  \nu^{\rm sgs}\frac{\partial^2
  \bar{u}}{\partial x^2}
\end{equation}
Here, for ease of illustration, let $\nu^{sgs}$ be a constant, though
in the Smagorinsky model it is proportional to a measure of the
local strain rate tensor.  On taking the Fourier transform of
equation~\ref{1dlesSMG}, and Euler forward time-stepping, we can
write
$$
\hat{\bar{u}}(k,t+\Delta t) = \hat{\bar{u}}(k,t)
- \Delta t ({\rm i}k \hat{f})
-\Delta t\,\nu^{\rm sgs}k^2\hat{\bar{u}}(k,t).
$$
Combining the 1st and 3rd terms on the rhs is equivalent to applying a
filter with response function $\hat{G}_S = 1 - \Delta t\,\nu^{\rm
  sgs}k^2$ to the solution at $t$.  This filtering provides the SGS
modeling. Although high wavenumber content is damped more (increasing
as $k^2$), {\em all} content is damped.  This is not to be considered
a wrong model, because the damping of any fixed range of low
wavenumber content will reduce as the grid is refined, but is,
therefore, a less efficient model.  When the dynamic Smagorinsky SGS
model is applied, $\nu^{\rm sgs}$ is determined during the course of
the computation from the solution.  The calibration coefficient that
appears in the expression for $\nu^{\rm sgs}$ varies with position and
time.  When the mean eddy viscosity
$$
<\nu^{\rm sgs}_{\rm Dynamic\,  Smagorinsky}>\,\, \le
\nu^{\rm sgs}_{\rm Smagorinsky} 
$$
suffices for stable computations, one can think of the dynamic model as
applying a mean $\nu^{\rm sgs}$ that is less than the standard model,
and thereby offering a better solution due to the smaller damping of
large scales.

The explicit filtering method described above has been called an
implicit LES (ILES) method, perhaps, because the numerical method was
designed for accuracy and stability and not explicitly for
LES~\cite{Visbal2003}.  An earlier method called MILES had also been
characterized as an ILES.  \citet{Boris1990} had explained the
effectiveness of MILES by stating, ``monotone convection algorithms
designed for positivity and causality, in effect have a minimal LES
filter and matching subgrid model already built in. [This ensures]
efficient transfer of the residual subgrid motions, [...] off the
resolved grid with minimal contamination of the well-resolved scales
by the numerical filter.''  These features turn out to be requirements
that appear in the derivation of the explicit filtering method.  A
variety of experiences with MILES, including an historical account, are
available~\cite{Grinstein2007}.  One way to understand the success of
MILES is to recall that in the FCT algorithm, the anti-diffusive step
is constructed from the local solution, and limited, to ensure that no
new extrema are created.  The FCT algorithm was designed to obtain
higher-order flow fields, capturing shocks without oscillations.  It
proves effective as an algorithm for LES because it suppresses
oscillations that will appear as nonlinear terms generate content at
wavenumbers larger than the ones that can be represented on the chosen
grid.  In its treatment of the difficulty at the high wavenumber end,
the algorithm is also optimal because it is designed to just prevent
the appearance of new extrema based on the local state.  Away from
locations where the integration would not produce a new extremum,
there is no modification of the solution.  The implied filter is then
active in a high wavenumber spectral buffer.  Since MILES was found to
be useful for LES without any explicit SGS model terms, the basic
numerical method remained of relatively low order---FCT is 2nd-order
in space.  On discovering that a compact scheme with a high-order
filter delivers useful LES without adding SGS model terms,
\citet{Visbal2003} have termed their method an ILES also.  The
explicit filtering methods cited here no longer attempt to provide any
kind of dynamic, optimal filtering.  Attempts in this direction did
not reveal any significant benefit by changing the filter cutoff, or
by reducing the frequency of its application.  For secondary filtering
(to be discussed below), the solution was not found to have any
sensitive dependence on the secondary filter parameter.

\subsection{Secondary filtering}\label{secfil}

\citet{Adams1999} had added a low-order relaxation regularization term
to the differential equation for the LES field.  In \citet{Stolz1999},
this was briefly mentioned as the use of a secondary filter that
improved the solution, but results were not included pending further
investigation.  It has been added and discussed in detail
subsequently~\cite{Stolz2001a, Stolz2001b}.  When this term is added,
the model eqn.~\ref{adm} would be modified to read
\begin{equation} \label{admreg}
\frac{\partial \bar{u}}{\partial t} + G * \frac{\partial
  f(Q*\bar{u})}{\partial x} = -\chi (I-Q*G)*\bar{u}. 
\end{equation}
Here, $\chi$ is a free parameter.  \citet{Stolz2001a} found solutions
to have but a weak dependence on $\chi$.  Mean velocity profiles
showed very little difference as $\chi$ was changed by a factor of 8.
As stated in \citet{Stolz2001a}, the effect of adding this relaxation
term can be realized by integrating without the additional term and
filtering the field $\bar{u}$ with filter $QG$ every $1/(\chi \Delta
t)$ timesteps.  Or, that applying the filter $QG$ to field $\bar {u}$
every $m$ timesteps while integrating eqn.\ref{adm} is equivalent to
integrating eqn.~(\ref{admreg}) with $\chi = 1/(m \Delta t$).  If
$m=1$, relaxation regularization is realized by applying the resultant
filter E*E = G*Q*G*Q to the evolving field.  For flat filters of the
type shown in Figs.~\ref{frfa} or \ref{scfrf},
$E(\alpha_1)*E(\alpha_1)$ can be approximated by applying filter
$E(\alpha_2)$, with $\alpha_2$ slightly less than $\alpha_1$.  So, a
formal secondary filtering step is not indicated since the
distinguishable benefit that would accrue is not evident.

\subsection{Sub-filter-scale and sub-grid-scale effects}\label{sgseffect}

This explicit filtering method does not distinguish between
sub-filter-scale and sub-grid-scale effects.  Sub-filter scales are
those represented on the LES grid ($k < k_{\rm max}$), but which may
have been distorted when the primary filter $G$ was applied to obtain
the equation for the LES variable $\bar{u}$.  Deconvolution of
$\bar{u}$ provides the field $u^*$ which has no content in $k > k_{\rm
  max}$.  Sub-grid scales are $k > k_{\rm max}$.  The remainder ${\cal
  R}$ in eqn.~(\ref{1dles}) has sub-grid scale contributions as well.
Indeed, in eqn.~(\ref{1dles}) even the term $\partial
f(\bar{u})/\partial x$ on the l.h.s. has spectral content in $k >
k_{\rm max}$ because it is nonlinear.  \citet{Winckelmans2001} have
examined modelling sub-filter and sub-grid scale effects.  Here, this
distinction is not made.  The modeling of the sub-filter effect is
obtained implicitly by using numerical schemes with flat filtering
characteristics below a cut-off $k_c < k_{\rm max}$, and the spectral
buffer provides the `sub-grid' scale modeling for {\em all} $k > k_c$.
We may expect this to suffice from the following argument: Consider
spectral components of the solution $\bar{u}$ at wavenumbers $k_l$,
$k_c$ and $k_s$.  $k_l \ll k_c$ is a large scale, and $k_s > k_c$ is a
small scale.  Quadratic difference-interactions among small scales of
amplitude $A_s$ can add $O(A_s^2)$ to the large scale.  This
contribution is small when the spectrum decays as in turbulent flows
{\em and} amplitudes near cut-off $A_c \ll A_l$, the amplitude of the
large scale component.  In LES with explicit filtering, it is
necessary for the cutoff wavenumber to lie (somewhere) in the inertial
range, {\em and} for some grid refinement to ensure that there are no
further, significant, qualitative changes to the solution.
Interactions between $k_c$ and $k_s$ are, similarly, not significant.
Quadratic sum interactions among small scales do not contribute to
large scales and do not represent a sub-grid-scale effect in an LES.
Interactions between scales of order $k_l$ and $k_s$ are not expected
to be significant either.  Within the range $k < k_c$, backscatter is
captured accurately, and imperfectly in $k_c < k < k_{\rm max}$.
Backscatter from scales $k > k_{\rm max}$ is not captured, but, as
explained above, it is not significant.

\section{Conclusions}

The explicit filtering method for LES comprises integration of the
governing equations without any added SGS terms and the application of
a flat low-pass filter to the transported fields after every
integration step.  The effective spatial filtering of several such
LES, including the filtering implied by the spatial operations of the
numerical schemes, was examined.  A common feature of these
implementations is a spectral buffer over a small part of the high
wavenumber end of the range of represented scales, analogous to buffer
or sponge zones near outflow boundaries.  Since there is essentially
no filtering of a range of large scales, and, as expected for LES, the
smallest represented scales are in the inertial range where the
amplitudes are small, when the scale range is increased, solutions
converge monotonically to the full spectrum (DNS), {\em without} any
significant changes to the large scale parts.  The monotonic
convergence of gross quantities (means and low order moments) is a
consequence of adding only at the high wavenumber end as the grid is
refined. Although it is not surprising that the {\em procedure} has
been seen as an example of an implicit LES (ILES), as a clean-up
operation, or, as a numerical operation to suppress (undefined)
instabilities, it ought to be clear from the discussion above that the
explicit filtering method provides a model for obtaining an LES.  The
principle revealed herein is quite general and can be used to
understand the observed or potential effectiveness of other methods
for LES as well.

\section*{Acknowledgments}
I thank my students Dr. S Chakravorty and Mr. Sumit Patel who provided
data from their simulations for the discussions in \S~\ref{sectrj}.  I
thank the reviewer who pointed out a prior use of the term `spectral
buffer' by Adams~\cite{Adams2011}.

\section*{References}
\bibliography{refs}

\end{document}